\documentclass[%
notitlepage,
twocolumn,
nolongbibliography,
 amsmath,amssymb,
]{revtex4-2}

\usepackage{graphicx}
\usepackage{dcolumn}
\usepackage{bm}
\usepackage{float}
\usepackage{units}
\usepackage{ulem}		

\usepackage{xcolor}
\usepackage{mathtools}
\usepackage{bbold}
\usepackage[]{graphicx}
\usepackage{tabularx}
\usepackage{hyperref}

\hypersetup{colorlinks=true, citecolor=blue, urlcolor=blue, linkcolor=blue}

\usepackage{empheq}

\usepackage[most]{tcolorbox}

\tcbset{colback=yellow!10!white, colframe=black!50!black, 
        highlight math style= {enhanced, 
            colframe=black,colback=red!10!white,boxsep=0pt}
        }

\newcommand{\kk}{\mathbf{k}}
\newcommand{\KK}{\mathbf{K}}
\newcommand{\QQ}{\mathbf{Q}}
\newcommand{\qq}{\mathbf{q}}

\definecolor{amber}{rgb}{0.82, 0.1, 0.26}

\begin{document}
\normalsize
\preprint{APS/123-QED}
\title{Fermionic vs. bosonic thermalization in the phonon-driven exciton dynamics: An analytic dimensionality study}

\author{Manuel Katzer}
\email{manuel.katzer@physik.tu-berlin.de}
\author{Malte Selig}
\author{Andreas Knorr}
\affiliation{Technische Universit\"at Berlin, Institut f\"ur Theoretische Physik, Nichtlineare Optik und Quantenelektronik, Hardenbergstra{\ss}e 36, 10623 Berlin, Germany}

\date{\today}

\begin{abstract}
Excitons are compound particles formed from an electron and a hole in semiconductors. The impact of this substructure on the phonon-exciton interaction is described by a closed system of microscopic scattering equations. To calculate the actual excitonic thermalization properties beyond the pure bosonic picture, this equation is derived directly from an electron-hole picture within the Heisenberg equation of motion framework. In addition to the well-known bosonic character of the compound particles, we identified processes of a repulsive, fermionic type, as well as attractive carrier exchange contributing to the scattering process. In this analytical study we give general statements about the thermalization of excitons in two and three dimensional semiconductors. We give insights on the strong dependence of the thermalization characteristics of the exciton Bohr radius and the thermalization wavelength. Above all, we analytically provide arguments why a bosonic behavior of excitons - such as an enhanced ground state occupation - requires the dominant phonon scattering to be quasielastic. Acoustic phonons tend to fulfil this, as each scattering event only takes small amounts of energy out of the distribution, while optical phonons tend to prevent macroscopic occupations of the lowest exciton state, since the Pauli repulsion between the individual carriers will then dominate the thermalization dynamics.

%
\end{abstract}





\maketitle

\section{Introduction}
Van der Waals heterostructures of atomically thin semiconductors sparked new hope to find bosonic or even macroscopic occupation effects of excitons, since they can host long living excitonic interlayer states~\cite{rivera2015observation,miller2017longlived,wang2019evidence,sigl2020signatures,sigl2022optical,troue2023extended,lohof2023confinedstate}. The discussion whether semiconductor excitons can show macroscopic occupation and spontaneous emergence of coherence dates back more than half a century~\cite{blatt1962boseeinstein,moskalenko1962inverse,keldysh1965possible,snoke2002spontaneous}, and there are reports of experimental signatures of related effects also for excitons in other semiconductor platforms, e.g., in GaAs Quantum Wells~\cite{high2012condensation,alloing2014evidence}, in quantum hall systems~\cite{eisenstein2014exciton}, and recently in bulk Cu$_2$O~\cite{morita2022observation}.
The last decades have also seen quite a few theoretical approaches towards effects of macroscopic occupation and spontaneous coherence in excitonic systems, and even more on exciton-polaritons. There is, e.g., the quantum kinetic approach from the Haug group~\cite{schmitt1999exciton,banyai2000condensation,schmitt2001boseeinstein,deng2010exciton}, and also an abundance of other theory works, e.g., Refs.~\cite{tassone1999excitonexciton,porras2002polariton,sarchi2007longrange,wouters2007excitations,ma2020spiraling,lagoin2021key,luders2021quantifying,cataldini2022emergent,remez2022leaky,luders2023tracking}. They are very diverse in their theoretical approaches, however, to our understanding, they all implicitly or explicitly share one key assumption, namely that excitons are pure bosons also for densities beyond the classical Maxwell-Boltzmann limit, which is a necessary condition to apply Bogoliubov approximations or Gross-Pittaevskii approaches~\cite{pitaevskii2016boseeinstein}. It was, however, also shown in several works that the fermionic substructure of excitons cannot be neglected at elevated densities~\cite{ivanov1993selfconsistent,steinhoff2017exciton,katsch2018theory,katsch2020excitonscatteringinduced,katsch2020optical,trovatello2022disentangling}. In a recent numerical study~\cite{katzer2023excitonphonon}, we challenged the assumption of pure bosonic thermalization, presenting an excitonic Boltzmann scattering equation to account for phonon mediated excitonic thermalization above the classical Maxwell-Boltzmann limit, taking for the first time the fermionic substructure into account. We showed that for large parts of the parameter space, fermionic Pauli blocking inhibits bosonic thermalization and thus resulting effects such as macroscopic ground state occupations. The equations of motion we study here, as derived and numerically approached in Ref.~\cite{katzer2023excitonphonon} are, in principle, valid for excitons of arbitrary dimension in semiconductors. 
In the present study, we provide analytic limits of a generalized exciton-phonon interaction dynamics, allowing us to deduce statements on the nature of the exciton as a particle between boson and fermion in a more general framework, and predict its behavior in two and three dimensions and in a unitless and therefore material insensitive description. We show that three parameters influence the thermalization, namely the exciton Bohr radius $a_0$, the thermal de Broglie wavelength $\lambda_{th}$, and their value relative to the phonon momentum $Q_\text{phon}$ which couples to the excitonic ground mode $\QQ=0$. The two main findings are: The Bohr radius of the exciton needs to be significantly smaller than the thermal wavelength, and only when the dominant exciton-phonon scattering process is elastic enough, stimulated scattering to the ground state can win over the Pauli repulsion between the carriers which consititute the excitons, as inelastic, optical phonon scattering favors Pauli blocking over stimulated scattering.

The paper is structured as follows: In Sec.~\ref{Sec:equationsection} we present the excitonic Boltzmann equation which was derived in Ref.~\cite{katzer2023excitonphonon}, and the individual contributions within the equation are recapitulated. In Sec.~\ref{Sec:approximativederivation} we give a detailed discussion on the analytic limit of low temperatures and derive analytic expressions allowing to interpret the behavior of excitons at low temperatures at the threshold of the first deviation from the classical Maxwell-Boltzmann limit. In Sec.~\ref{Sec:analyticresults} we give visualizations of the analytic expressions and discuss the resulting requirements for bosonic or fermionic thermalization behavior of excitons in two or three dimensions. In Sec.~\ref{Sec:Conclusion} we conclude.
\section{Excitonic Boltzmann equation}\label{Sec:equationsection}
This section is a brief recapitulation of the equation that was derived and introduced in Ref.~\cite{katzer2023excitonphonon}, before we examine its analytical limits in detail in the subsequent sections. The kinetic equation describes the dynamics of the exciton occupation,
\begin{align}
N_\QQ^\nu = \sum_{\qq\qq'} (\varphi_\qq^\nu)^* \varphi_{\qq'}^\nu \langle v_{\qq+\tilde\alpha\QQ}^\dagger c_{\qq-\tilde\beta\QQ} c_{\qq'-\tilde\beta\QQ}^\dagger v_{\qq'+\tilde\alpha\QQ}\rangle^c,
\end{align}
with center-of-mass momentum~$\QQ$ and the excitonic Rydberg state $\nu$, where $\varphi_\qq^\nu$ accounts for the real space relative motion wavefunction of the exciton gained from solving the Wannier equation~\cite{berghauser2014analytical}. The wavefunctions represent a full orthonormalized set $\sum_\qq (\varphi_\qq^{\lambda})^*\varphi_\qq^{\nu}=\delta^{\lambda\nu}$, while $c_\qq^{(\dagger)}$ and $v_\qq^{(\dagger)}$ are the fermionic anihilation (creation) operators for carriers in the conduction band and the valence band, respectively. The relative electron and hole masses $\tilde\alpha=\frac{m_e}{M}$ and $\tilde\beta=\frac{m_h}{M}$ abbreviate the respective proportion of the exciton mass $M=m_e+m_h$. 
It is important to state that for the derivation of the equation of motion for $N_\QQ^\nu$, in Ref.~\cite{katzer2023excitonphonon} we started from the fundamental electronic semiconductor Hamiltonian~\cite{haug2009quantum_short}, which allowed to account for bosonic and fermionic properties of the exciton thermalization:

\begin{align}\label{eq:hamiltonian}
    H
    &=
    \sum_{\kk\lambda}
    \epsilon_\kk^{\lambda}
    \lambda_\kk^{\dagger}
    \lambda_\kk
    +
    \sum_{\qq\alpha}
    \hbar \omega_\qq^\alpha
    b_\qq^{\dagger\alpha}
    b_\qq^\alpha
    \nonumber\\
    &\qquad
    +
    \frac{1}{2}
    \sum_{\lambda\lambda'\kk\kk'\qq}
    V_\qq
    \lambda_{\kk}^\dagger
    {\lambda'}_{\kk'}^\dagger
    {\lambda'}_{\kk'+\qq}
    \lambda_{\kk-\qq}
    \nonumber\\
    &\qquad
    +
    \sum_{\kk\qq\lambda\alpha}
    g_\qq^{\lambda\alpha}
    \lambda_{\kk+\qq}^\dagger
    \lambda_\kk
    (b^\alpha_\qq+b_{-\qq}^{\dagger\alpha}).
\end{align}

The first term accounts for the dispersion of electrons in the conduction band ($\lambda = c$) and valence band ($\lambda = v$) and momentum $\mathbf{k}$ being parametrized from DFT calculations in effective mass approximation~\cite{kormanyos2015theory}.  The second term accounts for the dispersion of phonons. The mode index $\alpha$ accounts for acoustic and optical phonon modes, parametrized by \textit{ab initio} values from the literature, for TMDC excitons see e.g.~Refs.~\cite{kaasbjerg2012phononlimited,li2013intrinsic,jin2014intrinsic}. The third term accounts for the Coulomb interaction between electrons and holes. The coupling element $V_\qq$ is obtained from an analytic solution of the Poisson equation for the Rytova-Keldysh potential~\cite{rytova1967screened,rytova2020screened}. The fourth term accounts for the electron phonon interaction in valence and conduction band. The appearing electron-phonon coupling elements $g_{\qq}^{\lambda\alpha}$, for the different involved phonon modes $\alpha$ in the two bands $\lambda=c,v$ are treated in effective deformation potential approximation, parametrized with values typically obtained from DFT calculations, see e.g. for TMDCs Refs.~\cite{li2013intrinsic,jin2014intrinsic,selig2016excitonic,selig2018dark,selig2020suppression}.

The excitonic Boltzmann scattering equation above the Maxwell-Boltzmann limit was obtained by treating the appearing hierarchy problem in Born-Markov approximation and applying the unit operator technique~\cite{katsch2018theory,katsch2020excitonscatteringinduced,ivanov1993selfconsistent}, which allows to project the fermionic expectation operators on excitonic pair occupation operators. The strict calculation in the electron-hole-picture also circumvents difficulties in the factorization of excitonic expectation values. For details on the derivation see Ref.~\cite{katzer2023excitonphonon}. 
\begin{figure}[t!]
    \centering
    \includegraphics[width=\linewidth]{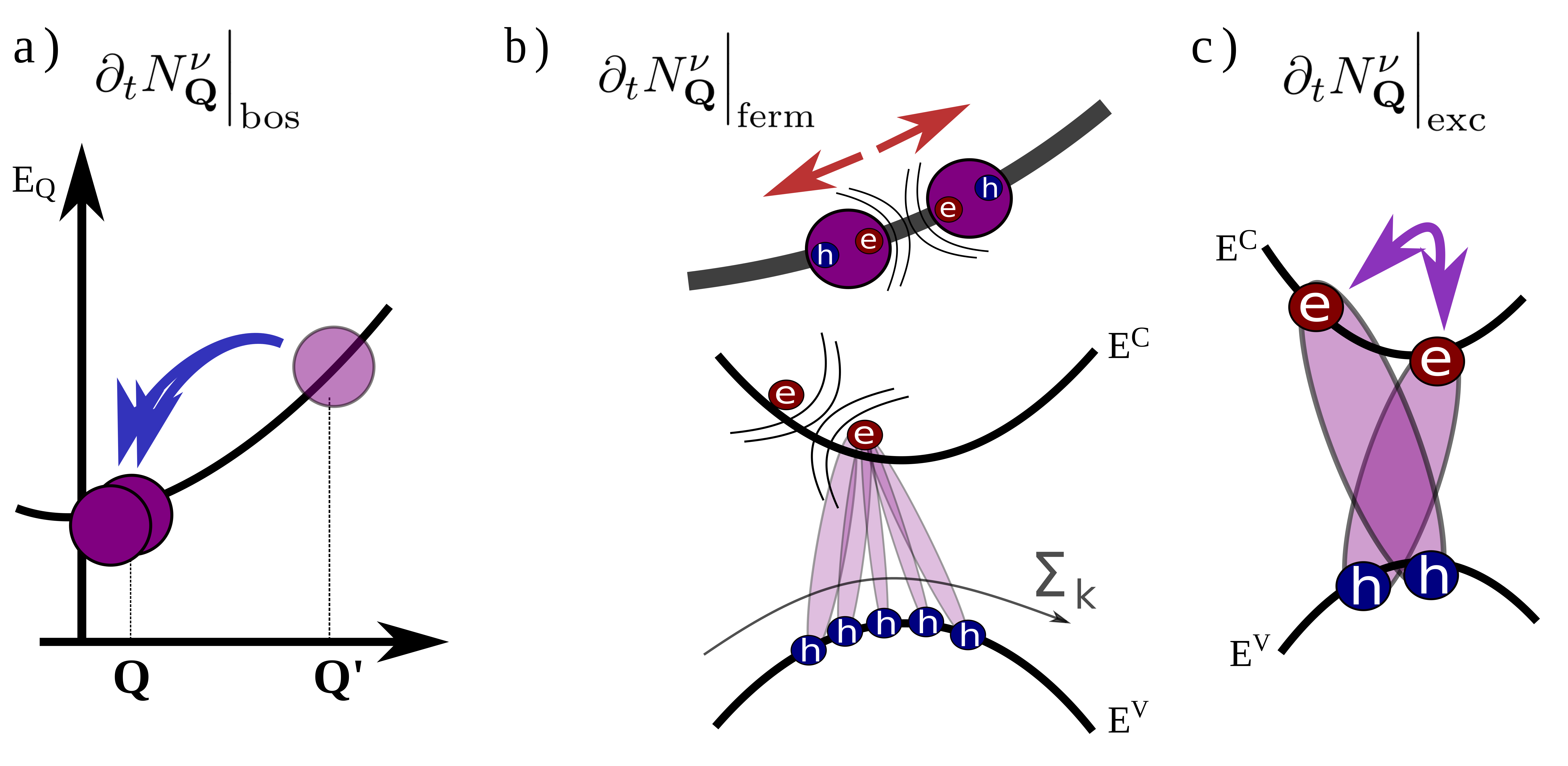}
    \caption{
            Recapitulation of the different nonlinear effects in the excitonic thermalization process, as introduced in detail in Ref.~\cite{katzer2023excitonphonon}. a) The nonlinearity $\partial_t N_\QQ \big|_\text{bos}$ leads to stimulated scattering, similar to pure bosonic particles. b) The fermionic correction term $\partial_t N_\QQ \big|_\text{ferm}$ leads to a repulsion as electrons (and holes) show Pauli blocking. Many excitons contribute to this effect at a given momentum, illustrated by the summation over k in the lower panel. c) The exchange-nonlinearity, $\partial_t N_\QQ \big|_\text{exc}$, is of attractive nature. It is due to a carrier exchange during the scattering process. Figure adapted from~\cite{katzer2023excitonphonon}.}
    \label{fig:sketch}
\end{figure}
The resulting equation reads

\begin{align}\label{eq:mainequation}
    &\partial_t
    N_{\QQ}^{\nu}
    =
    \partial_t
    N_{\QQ}^{\nu}
    \Big|_{\text{class}}
    +
    \partial_t
    N_{\QQ}^{\nu}
    \Big|_{\text{bos}}
    +
    \partial_t
    N_{\QQ}^{\nu}
    \Big|_{\text{ferm}}
    +
    \partial_t
    N_{\QQ}^{\nu}
    \Big|_{\text{exc}}.
\end{align}

The first term in Eq.~(\ref{eq:mainequation}) accounts for the linear contribution, responsible for a thermalization according to the classical Maxwell-Boltzmann statistics. It is valid for dilute, classical exciton gases~\cite{selig2019ultrafast,selig2020suppression,katzer2023impactnessy}, and reads
\begin{align}\label{eq:classicboltzmann}
    \partial_t
    N_{\QQ}^{\nu}
    \Big|_{\text{class}}
    &=
    \frac{2\pi}{\hbar}
    \sum_{\QQ'\lambda}
    \Big(
    W_{\QQ'\QQ}^{\lambda\nu}
    N_{\QQ'}^\lambda
    -
    W_{\QQ\QQ'}^{\nu\lambda}
    N_{\QQ}^\nu
    \Big),
\end{align}
with the scattering tensor
\begin{align}\label{eq:linscatmat}
    &W_{\QQ\QQ'}^{\nu\lambda}
    \nonumber\\
    &=
    \sum_\alpha
    |g_{|\QQ-\QQ'|,\alpha}^{\nu\lambda}|^2
    \Big(
    (1+n_{|\QQ-\QQ'|}^\alpha)
    \delta
    \big(
    \mathcal{E}_{\QQ'}^{\lambda}
	-
	\mathcal{E}_{\QQ}^{\nu}
	+
	\hbar\omega_{|\QQ-\QQ'|}^\alpha
	\big)\nonumber\\
	&\qquad\qquad
	+
    n_{|\QQ-\QQ'|}^\alpha
    \delta
    \big(
    \mathcal{E}_{\QQ'}^{\lambda}
	-
	\mathcal{E}_{\QQ}^{\nu}
	-
	\hbar\omega_{|\QQ'-\QQ|}\alpha
	\big)
	\Big).
\end{align}
The exciton-phonon coupling reads 

\begin{align}\label{eq:standard_excitonic_g}
g_{\QQ,\alpha}^{\nu\nu'}
=
\sum_\qq
\Big(
g_{\QQ}^{c\alpha}
(\varphi_\qq^\nu)^*
\varphi_{\qq-\tilde\beta\QQ}^{\nu'}
-
g_{\QQ}^{v\alpha}
(\varphi_\qq^\nu)^*
\varphi_{\qq+\tilde\alpha\QQ}^{\nu'}
\Big),
\end{align}
and depends on convolutions of exciton wavefunctions $\varphi_\qq^\nu$ and the electron-phonon coupling element $g_{\qq}^{\lambda\alpha}$ from the electronic Hamiltonian, Eq.~(\ref{eq:hamiltonian}).
For densities above the classical limit, additionally three nonlinearities become important, all in the same order of the exciton density 
\begin{align}\label{eq:densitydefinition}
\bar n = \frac{1}{L^d}\sum_{\QQ\nu}N_\QQ^\nu,
\end{align}
 with a factor $L^d$ of unit $[L^d]=\unit[]{nm^{d}}$, i.e., an area or a volume, dependent on the dimension $d\in\{2,3\}$, which we introduce here as an index to allow for a more general formulation throughout the whole manuscript. The nonlinearites are all sketched in Fig.~\ref{fig:sketch}. The second term in Eq.~(\ref{eq:mainequation}) accounts for bosonic stimulated scattering, which would also occur for pure bosons~\cite{banyai2000condensation}, cp.~Fig.~\ref{fig:sketch}(a)
\begin{align}\label{eq:bosonicNL}
    \partial_t
    N_{\QQ}^{\nu}
    \Big|_{\text{bos}}
    &=
    \frac{2\pi}{\hbar}
    \sum_{\QQ'\lambda}
    \Gamma_{\QQ\QQ'}^{B,\nu\lambda}
    N_{\QQ'}^{\lambda} 
    N_{\QQ}^\nu.
\end{align}
The respective scattering matrix can be directly computed from the classical scattering matrix $W_{\QQ\QQ'}^{\nu\lambda}$, Eq.~(\ref{eq:linscatmat}), and reads
\begin{align}\label{eq:bosscatmat}
    \Gamma_{\QQ\QQ'}^{B,\nu\lambda}
    &=
    W_{\QQ\QQ'}^{\nu\lambda}
    -
    W_{\QQ'\QQ}^{\lambda\nu}
    \nonumber\\
    &
    =
    \sum_\alpha
    |g_{|\QQ-\QQ'|,\alpha}^{\nu\lambda}|^2
    \Big(
    \delta
    \big(
    \mathcal{E}_{\QQ'}^{\lambda}
	-
	\mathcal{E}_{\QQ}^{\nu}
	-
	\hbar\omega_{|\QQ-\QQ'|}^\alpha
	\big)
    \nonumber\\
	&\qquad\qquad\qquad
	-
    \delta
    \big(
    \mathcal{E}_{\QQ'}^{\lambda}
	-
	\mathcal{E}_{\QQ}^{\nu}
	+
	\hbar\omega_{|\QQ'-\QQ|}^\alpha
	\big)
	\Big).
\end{align}
The mentioned stimulated scattering of this bosonic nonlinear contribution, Eq.~(\ref{eq:bosonicNL}), leads to amplified scattering to already strongly occupied states, and thus enables high occupations of the ground state at $\QQ=0$. 

In addition to the classical and bosonic contributions to Eq.~(\ref{eq:mainequation}), two further nonlinearities occur due to the electron-hole substructure of the excitons. The third term in Eq.~(\ref{eq:mainequation}), cp.~Fig.~\ref{fig:sketch}(b), is of repulsive nature and occurs due to Pauli blocking of the fermionic carriers the exciton is constituted of. It reads

\begin{align}\label{eq:fermionicNL}
    \partial_t
    N_{\QQ}^{\nu}
    \Big|_{\text{ferm}}
    &=
    \frac{2\pi}{\hbar}
    \sum_{\QQ'\KK\lambda\nu'}
    \Big(
    \Gamma_{\QQ'\QQ,\KK}^{F,\nu\lambda,\nu'}
    N_{\QQ'}^\lambda
    -
    \Gamma_{\QQ\QQ',\KK}^{F,\lambda\nu,\nu'}
    N_{\QQ}^\nu
    \Big)
    N_{\KK}^{\nu'}.
\end{align}
Note that compared to classical and bosonic contributions of Eq.~(\ref{eq:mainequation}), this repulsive term requires an additional convolution over all excitonic states, which orginates from the projection into the excitonic basis. 
Similar to the scattering matrices above, we find here the three dimensional scattering tensor 
\begin{align}\label{eq:fermionicscatteringtensor}
    &\Gamma_{\QQ\QQ',\KK}^{F,\lambda\nu,\nu'}
    \nonumber\\
    &=
    \sum_\alpha
	Re(
	g_{\QQ-\QQ',\alpha}^{\nu\lambda}
	g_{\QQ'\QQ,\KK,\alpha}^{F,\lambda\nu,\nu'}
    )
    \Big(
    \delta
    \big(
    \mathcal{E}_{\QQ'}^{\lambda}
	-
	\mathcal{E}_{\QQ}^{\nu}
	-
	\hbar\omega_{\QQ-\QQ'}^\alpha
	\big)
	\nonumber\\
	&\qquad\qquad
	-
    \delta
    \big(
    \mathcal{E}_{\QQ'}^{\lambda}
	-
	\mathcal{E}_{\QQ}^{\nu}
	+
	\hbar\omega_{\QQ'-\QQ}^\alpha
	\big)
	\Big).
\end{align}
Furthermore, there occurs a second nonlinearity due to the electronic substructure of the exciton, namely the fourth term in Eq.~(\ref{eq:mainequation}), cp.~Fig.~\ref{fig:sketch}(c). It results from exchanging carriers between excitons during the scattering and constitutes a typical fermionic exchange nonlinearity:
\begin{align}
    \partial_t
    N_{\QQ}^{\nu}
    \Big|_{\text{exc}}
    &=
    \frac{2\pi}{\hbar}
    \sum_{\KK\KK'\lambda'\nu'}
    \Gamma_{\QQ,\KK,\KK'}^{E,\nu\lambda'\nu'}
    N_{\KK'}^{\lambda'}
    N_{\KK}^{\nu'}.
    \label{eq:exchangemain}
\end{align}
Here, the respective scattering tensor reads
\begin{align}\label{eq:exchangescatteringtensor}
    &\Gamma_{\QQ,\KK,\KK'}^{E,\nu\lambda'\nu'}
    \nonumber\\
    &=
    \frac{1}{2}
    \sum_{\QQ'\alpha\lambda}
    \Big(
    g_{\QQ-\QQ',\alpha}^{\nu\lambda}
    g_{\QQ'\QQ,\KK,\KK',\alpha}^{E,\nu\lambda,\lambda',\nu'}
    +
    g_{\QQ'-\QQ,\alpha}^{\lambda\nu}
    g_{\QQ\QQ',\KK,\KK',\alpha}^{E,\lambda\nu,\lambda',\nu'}
    \Big)
    \nonumber\\
    &\qquad
    \times
    \Big(
    \delta
    \big(
    \mathcal{E}_{\QQ'}^{\lambda}
	-
	\mathcal{E}_{\QQ}^{\nu}
	-
	\hbar\omega_{|\QQ-\QQ'|}^\alpha
	\big)
	\nonumber\\
	&\qquad\qquad\qquad
	-
    \delta
    \big(
    \mathcal{E}_{\QQ'}^{\lambda}
	-
	\mathcal{E}_{\QQ}^{\nu}
	+
	\hbar\omega_{|\QQ'-\QQ|}^\alpha
	\big)
	\Big).
\end{align}
For all contributions of Eq.~(\ref{eq:mainequation}), the coupling constants for the exciton-phonon-coupling are given by the wavefunction overlap of the involved carriers, which depend on the momenta $\QQ'$ of the phonons associated to the process. While for the classical an bosonic contribution, this is encoded in the well known coupling $g_{\QQ,\alpha}^{\nu\nu'}$, Eq.~(\ref{eq:standard_excitonic_g}), the new scattering tensors that arise due to the fermionic substructure come with new, more elaborate overlaps, namely $g_{\QQ'\QQ,\KK,\alpha}^{F,\lambda\nu,\nu'}$ for the fermionic nonlinearity, and $g_{\QQ\QQ',\KK,\KK',\alpha}^{E,\lambda\nu,\lambda',\nu'}$ for the exchange nonlinearity (see App.~\ref{app:overlapappendix} for details). Their increased complexity is reflecting the convolution with all carriers involved, (i.e., also those of the other excitons). The fact that the fermionic scattering tensors depend on more overlapping excitonic wavefunctions compared to the classical and bosonic terms, leads to a strong dependence of the relative dominance between the nonlinearities on the exciton Bohr radius $a_0$. Large Bohr radii lead to smaller wavefunctions in momentum space. This will be crucial for the analytic discussion in the following section.

\section{Analytical Limit}\label{Sec:approximativederivation}
Equation~(\ref{eq:mainequation}) is a general result which is valid for excitons in different systems, thus also for systems with different dimensionalitiy. The following discussion is thus conducted for two and three dimensional excitons, respectively, with TMDC excitons used to give an example of experimentally accessible parameter ranges in two dimensions. For our analytical discussion, we rewrite Eq.~(\ref{eq:mainequation}) to a more compact form:

\begin{align}\label{eq:mainequationalternative}
    &\partial_t
    N_{\QQ}^{\nu}
    \nonumber\\
    &=
    \frac{2\pi}{\hbar}
    \Bigg[
    \sum_{\QQ'\lambda}
    \bigg[
    \Big(
    W_{\QQ'\QQ}^{\lambda\nu}
    \big(1+N_{\QQ}^{\nu}\big)
    -
    \sum_{\KK\nu'}
    \Gamma_{\QQ'\QQ,\KK}^{F,\nu\lambda,\nu'}
    N_{\KK}^{\nu'}
    \Big)
    N_{\QQ'}^\lambda
    \nonumber\\
    &\qquad\qquad
    -
    \Big(
    W_{\QQ\QQ'}^{\nu\lambda}
    \big(1+ N_{\QQ'}^{\lambda}\big)
    -
    \sum_{\KK\nu'}
    \Gamma_{\QQ\QQ',\KK}^{F,\lambda\nu,\nu'}
    N_{\KK}^{\nu'}
    \Big)
    N_{\QQ}^\nu
    \bigg]
    \nonumber\\
    &\qquad\qquad
    -
    \sum_{\KK\KK'\lambda'\nu'}
    \Gamma_{\QQ,\KK,\KK'}^{E,\nu\lambda'\nu'}
    N_{\KK'}^{\lambda'}
    N_{\KK}^{\nu'}
    \Bigg].
\end{align}

In order to identify parameter regimes where the thermalization is dominated by the bosonic nonlinearities, we introduce two general parameters, the Bohr radius $a_0$ and the thermal wave length $\lambda_{th}$. To formally define an exciton Bohr radius, the 1s wave functions, which are accessed as the eigenfunctions of the Wannier equation, can be fitted to an analytical model~\cite{haug2009quantum_short}, which in the two- and three-dimensional case read:
\begin{align}\label{eq:bohrradiusmodel}
    &\varphi_\qq^{d=2}
    =
    \frac{8\sqrt{2\pi a_0^2/L^2}}{(4+a_0^2q^2)^\frac{3}{2}}
    &\varphi_\qq^{d=3}
    =
    \frac{8\sqrt{\pi a_0^3 /L^3}}
    {(1+a_0^2q^2)^2}
\end{align}
In the three-dimensional model, $a_0$ is directly equivalent to the extension of the exciton in real space, while for two dimensions, it is typically defined to be twice the radius of the extension~\cite{haug2009quantum_short}. TMDC excitons, e.g., typically have extension radii in the order of $\unit[1]{nm}$~\cite{berkelbach2013theory}, we therefore estimate the Bohr radius for TMDCs to be around $\unit[2]{nm}$, when we give examples in the following. Furthermore, we introduce the thermal wavelength $\lambda_{th}$ (also referred to as de Broglie wavelength), which depends on temperature $T$ and effective exciton mass $M$:
\begin{align}
    \lambda_{th} = \frac{\hbar}{\sqrt{2Mk_BT}}
\end{align}
The thermal wavelength $\lambda_{th}$ characterizes a typical inverse wavelength extension of the occupation number distribution as a function of wave numbers in an ideal classical exciton gas at a specified temperature and thus is a well defined parameter close to the classical limit~\cite{haug2008quantum_short}. In the following, we discuss analytical limits of Eq.~(\ref{eq:mainequation}).

\begin{figure}[t!]
    \centering
    \includegraphics[width=\linewidth]{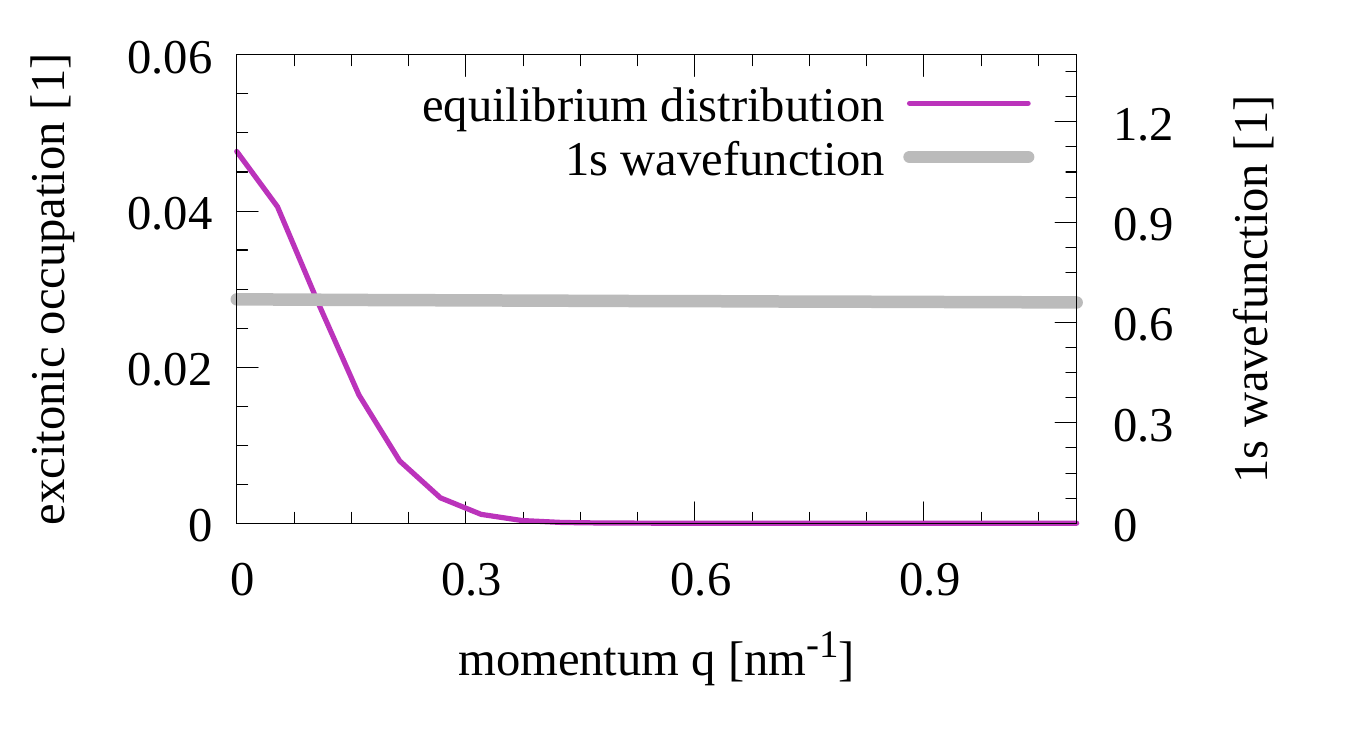}
    \caption{Equilibrium distribution $N_\QQ$ @ 10K for a Bohr radius $a_0 \approx \unit[0.2]{nm}$ in momentum space when computed with the full equation (\ref{eq:mainequation}) exemplarily for TMDC parameters, in comparison to the respective 1s wavefunction $\varphi_\qq$. The wavefunction can in very good approximation be estimated as flat in the momentum region of the exciton dynamics.}
    \label{fig:wavefunction_flat}
\end{figure}

\subsection{Low temperature limit}\label{sec:lowtempsubsection}
In the following analytic discussion of Eq.~(\ref{eq:mainequation}), we assume that the Bohr radius is small compared to the thermal wavelength:
\begin{align}\label{eq:approxablam}
    a_0 \ll \lambda_{th}, 
\end{align}
i.e., the average particle wavelength is large compared to its Bohr radius. We can also express this w.r.t.\ temperature, and assume $M\approx 1.1m_{el}$ and $a_0\approx 2nm$ (typical TMDC values) to get an idea for which temperatures this approximation is valid e.g. for TMDC excitons:
\begin{align}\label{eq:approxablamT}
    T \ll  \frac{\hbar^2}{2Mk_Ba_0^2} \approx  98K
\end{align}
This shows that also for experimentally accessible regimes for instance in TMDCs, at low temperatures, the analyzed limit remains a good approximation over a large parameter range. In addition, translated to momentum space, this approximation implies that the wavefunction can be considered flat on the momentum scale of the thermalization dynamics, cp. Fig.~\ref{fig:wavefunction_flat}. This allows to approximate $\varphi_\qq  \approx \varphi_{\qq+\QQ'}$ in the appearing overlap integrals in the scattering matrices in Eqs.~(\ref{eq:standard_excitonic_g},\ref{eq:fermionicoverlaps},\ref{eq:exchangeoverlaps}), i.e., we set
\begin{align}\label{eq:intergral1approx}
\sum_\qq
(\varphi_\qq)^*
\varphi_{\qq-\beta\QQ'}
\approx 1,
\end{align}
and approximate other overlap integrals accordingly. In physical terms, this means that we neglect that the overlap integrals are slightly smaller than 1, as this is a small effect in this regime. We carefully checked that in the limit of Eq.~(\ref{eq:approxablam}), the full numerics~\cite{katzer2023excitonphonon} give very similar results with and without this assumption. If we apply this approximation to all scattering tensors, we can give analytic expressions depending on $a_0$ for the overlaps. This allows us to significantly simplify the main equation to
\begin{align}\label{eq:mainequationapprox}
    &\partial_t
    N_{\QQ}
    \nonumber\\
    &\approx
    \frac{2\pi}{\hbar}
    \sum_{\QQ'\alpha}
    \big|
    g_{\QQ'-\QQ}^{c\alpha}
    -
    g_{\QQ'-\QQ}^{v\alpha}
    \big|^2\nonumber\\
    &
    \times
    \bigg[
    \Big(\frac{1}{2}\pm \frac{1}{2} + n_{|\QQ-\QQ'|}^\alpha\Big)
    \delta
    \big(
    \mathcal{E}_{\QQ'}
	-
	\mathcal{E}_{\QQ}
	\pm
	\hbar\omega_{|\QQ-\QQ'|}^\alpha
	\big)
	N_{\QQ'}
    \nonumber\\
    &\qquad
    -
    \Big(\frac{1}{2}\pm \frac{1}{2} + n_{|\QQ-\QQ'|}^\alpha\Big)
    \delta
    \big(
    \mathcal{E}_{\QQ'}
	-
	\mathcal{E}_{\QQ}
	\mp
	\hbar\omega_{|\QQ-\QQ'|}^\alpha
	\big)
	N_{\QQ}
    \nonumber\\
    &
    +
    \Big(
    N_{\QQ'}
    N_{\QQ}
    -
    2
    \bar n
    a_0^d
    \mathcal{F}^d
    \big(
    N_{\QQ'}
    +
    N_{\QQ}
    \big)
    +
    (\bar n
    a_0^d)^2
    \mathcal{B}^d
    \Big)
    \nonumber\\
    &\qquad
    \times
    \Big(
    \delta
    \big(
    \mathcal{E}_{\QQ}
	-
	\mathcal{E}_{\QQ'}
	+
	\hbar\omega_{|\QQ-\QQ'|}^\alpha
	\big)
    \nonumber\\
    &\qquad\qquad\qquad
	-
    \delta
    \big(
    \mathcal{E}_{\QQ}
	-
	\mathcal{E}_{\QQ'}
	-
	\hbar\omega_{|\QQ'-\QQ|}^\alpha
	\big)
	\Big)
	\bigg].
\end{align}
Note that we identified the exciton density $\bar n$, Eq.~(\ref{eq:densitydefinition}). We have introduced dimension-dependent abbreviations $\mathcal{F}^d,\mathcal{B}^d, d\in\{2,3\}$, see App.~\ref{app:wavefunctionintegrals}, that allow us to give a dimension invariant derivation, making use of the fact that after the approximation, Eq.~(\ref{eq:intergral1approx}), the integrals over the wavefunction give only an analytically computable factor and a dependence of the Bohrradius $a_0$, for the fermionic term $a_0^d$, and for the exchange term even $(a_0^d)^2$. 
Note that this approximate form, Eq.~(\ref{eq:mainequationapprox}), is - as the full equation, Eq.~(\ref{eq:mainequation}) - density conserving, which can be seen when executing the sum over all momenta ($\sum_\QQ \partial_t N_\QQ $): For the nonlinearities, the density is conserved for each term separately, simply because the sum $\sum_{\QQ\QQ'\alpha}    \delta
    \big(
    \mathcal{E}_{\QQ}
	-
	\mathcal{E}_{\QQ'}
	+
	\hbar\omega_{|\QQ-\QQ'|}^\alpha
	\big)
	-
    \delta
    \big(
    \mathcal{E}_{\QQ}
	-
	\mathcal{E}_{\QQ'}
	-
	\hbar\omega_{|\QQ'-\QQ|}^\alpha
	\big)=0$.
 Eq.~(\ref{eq:mainequationapprox}) allows for a better understanding of the derived full excitonic scattering equation, Eq.~(\ref{eq:mainequation}): First of all, the first two lines represent the classical, linear case, which at approaching equilibrium drive the exciton distribution into the classical Maxwell-Boltzmann distribution. The temperature enters here directly via the phononic occupation number $n_\QQ^\alpha(T)$, i.e., the phonon equilibrium Bose distribution, thus the first two lines of Eq.~(\ref{eq:mainequationapprox}) correspond to the classical thermalization dynamics in dilute gases~\cite{selig2018dark,selig2019ultrafast,katzer2023impactnessy}. Above the low density limit, terms of the order of $\bar n^2$ become relevant, leading to a deviation from the Maxwell-Boltzmann distribution in equilibrium.
 The last three lines of Eq.~(\ref{eq:mainequationapprox}) account for these different nonlinearities induced by quantum effects beyond the classical gas dynamics. Evidently, all nonlinearities share the same phonon prefactor and energy-momentum selection rules. The first term, scaling as $N_{\QQ'}N_\QQ$, corresponds to the ideal, bosonic case and is independent of the unitless parameter $\eta=\bar n a_0^d$ (it is dependent on the density $\bar n$ via the square of the occupation, but not on the Bohr radius, as we will see in the following). In contrast, the corrections due to the fermionic substructure of the excitons are typically scaling in orders of $\bar n a_0^d$~\cite{katsch2018theory,ivanov1993selfconsistent}, and it thus is intuitive that also in the case of Eq.~(\ref{eq:mainequation}) those terms depend on this unitless parameter, the occuring nonlinearity with negative sign goes linear in $\eta$, the attractive exchange even with the square, $\eta^2$.

\begin{figure}
    \centering
    \includegraphics[width=\linewidth]{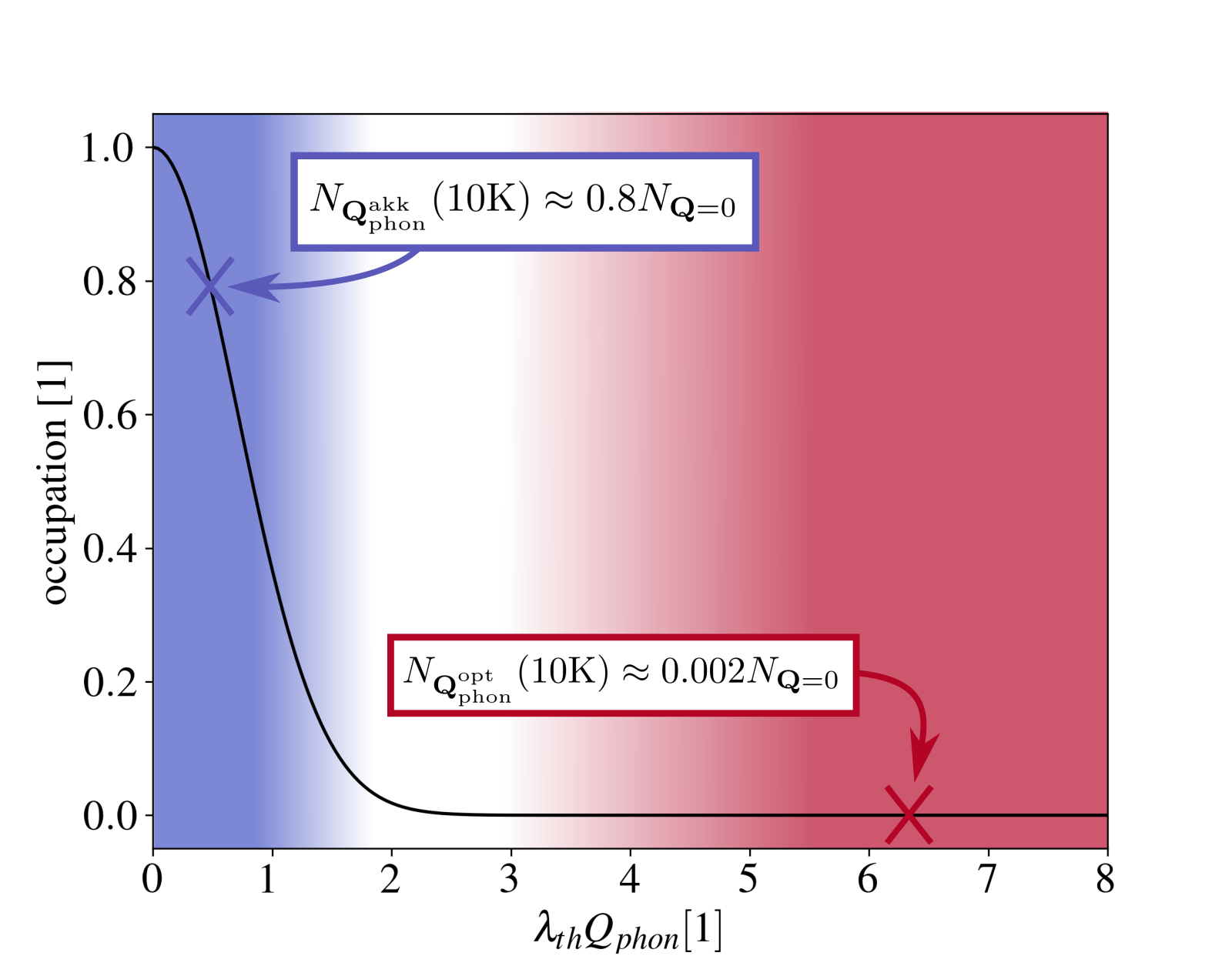}
    \caption{Visualization of the occupation $N_{\QQ'}$ of the scattering partner for the lowest $Q=0$ mode, compared to $N_{\QQ=0}$, cp. Eq.~(\ref{eq:mainequationapprox}), (but also in the main equation, Eq.~(\ref{eq:mainequation})). We show exemplary values at $T=\unit[10]{K}$, the lower the temperature, the lower also the occupation of the scattering partners. We can see here that for acoustic, rather elastic phonon scattering, the occupation of the scattering partner is high enough that the bosonic NL can win over the fermionic term in Eq.~(\ref{eq:mainequationapprox}) For optical scattering, the occupation of the exciton state that scatters to the ground mode is almost unoccupied. Thus, the repulsion will win over the stimulated scattering, as it is independent of this occupation.}
    \label{fig:boltzmann-sketch}
\end{figure}

Eq.~(\ref{eq:mainequationapprox}) also makes a fact visible which will become important in the following: The bosonic, stimulated scattering, i.e., the first nonlinear term, is dependent not only on the occupation $N_{\QQ}$ itself, but also on the occupation of the respective scattering partner $ N_{\QQ'}$, with the momentum $\QQ'$ determined by the Fermi selection rules (last two lines in Eq.~(\ref{eq:mainequationapprox})). For the stimulated scattering, especially the scattering to the ground state (the $\QQ=0$ mode) this is decisive: In the following we treat the scattering momentum provided by the phonons to fill the ground state as a key parameter for the classification of the exciton thermalization as of bosonic or fermionic tendency. We denote the respective momentum for scattering to the ground state $\QQ_\text{phon}$. Taking the phonon dispersion into account, for inelastic, optical phonon scattering, this momentum is comparatively large, typically in the range of $\QQ_\text{phon}\approx\unit[1]{nm^{-1}}$ for TMDC monolayers, while acoustic phonons show angular dependent, yet significantly smaller momenta in the range $\QQ_\text{phon}\approx\unit[0.05-0.1]{nm^{-1}}$~\cite{berkelbach2013theory,li2013intrinsic,jin2014intrinsic,selig2020suppression}. As a consequence, the occupation of the modes that provide the scattering to the ground state are significantly higher populated for acoustic phonons than for optical ones, as is illustrated for a monolayer MoSe$_2$ at $T=\unit[10]{K}$ in Fig.~\ref{fig:boltzmann-sketch}. This is important, since the fermionic Pauli repulsion  occurs independent of this specific occupation $N_{\QQ_\text{phon}}$, but sums over all occupations, and is thus independent of the scattering momentum, i.e., the second and third nonlinearities in Eq.~(\ref{eq:mainequation}), scaling with $\bar n a_0^d$ and $(\bar n a_0^d)^2$, do not rely on high values of the occupation $ N_{\QQ_\text{phon}}$ compared to the occupation of the ground mode, $ N_{\QQ=0}$. This is the reason why for dominant optical, inelastic phonon scattering, the excitonic thermalization cannot be bosonic, as we will show in more detail in the following.
\subsection{Small deviations from classical Maxwell-Boltzmann limit}
In the Boltzmann limit, the occupations for a given exicton density $\bar{n}$ read for arbitrary dimension $d$
\begin{align}\label{eq:boltzmanndistribution}
    N_\QQ
    ^\text{MB}
    \approx
    (2\pi^\frac{1}{2}\lambda_{th})^d
    \bar n
    e^{-\lambda_{th}^2Q^2}.
\end{align}
We hence look at small deviations $\delta  N_\QQ = N_\QQ - N_\QQ^\text{MB}$ from the occupation $ N_\QQ$. 
The leading order is identified, with
 $1\gg N_\QQ \gg N_\QQ N_\QQ \gg \delta N_\QQ \gg N_\QQ \delta N_\QQ \gg \delta N_\QQ \delta N_\QQ $. 
This leaves us with
\begin{align}\label{eq:mainequationapproxwithdelta2}
    &\partial_t
    \big(\delta
    N_{\QQ}
    \big)
    \nonumber\\
    &=
    \frac{2\pi}{\hbar}
    \sum_{\QQ'\alpha\pm}
    \big|
    g_{\QQ'-\QQ}^{c\alpha}
    -
    g_{\QQ'-\QQ}^{v\alpha}
    \big|^2\nonumber\\
    &
    \times
    \bigg[
    \Big(\frac{1}{2}\pm \frac{1}{2} + n_{|\QQ-\QQ'|}^\alpha\Big)
    \delta
    \big(
    \mathcal{E}_{\QQ'}
	-
	\mathcal{E}_{\QQ}
	\pm
	\hbar\omega_{|\QQ-\QQ'|}^\alpha
	\big)
	\delta
    N_{\QQ'}    
    \nonumber\\
    &\qquad
    -
    \Big(\frac{1}{2}\pm \frac{1}{2} + n_{|\QQ-\QQ'|}^\alpha\Big)
    \delta
    \big(
    \mathcal{E}_{\QQ'}
	-
	\mathcal{E}_{\QQ}
	\mp
	\hbar\omega_{|\QQ-\QQ'|}^\alpha
	\big)
	  \delta    
    N_{\QQ}    
    \nonumber\\
    &
    +
    \Big(    
    N_{\QQ'}
    ^\text{MB}    
    N_{\QQ}
    ^\text{MB}
    \nonumber\\
    &\qquad\qquad
    -
    2
    \bar n
    a_0^d
    \mathcal{F}^d
    \big[
    \big(
    N_{\QQ'}
    ^\text{MB}
    \big)
    +
    \big(    
    N_{\QQ}
    ^\text{MB}
    \big)
    \big]
    \nonumber\\
    &\qquad\qquad
    +
    (\bar n
    a_0^d)^2
    \mathcal{B}^d 
    \Big)
    \nonumber\\
    &\qquad
    \times
    \Big(
    \delta
    \big(
    \mathcal{E}_{\QQ}
	-
	\mathcal{E}_{\QQ'}
	+
	\hbar\omega_{|\QQ-\QQ'|}^\alpha
	\big)\nonumber\\
    &\qquad\qquad\qquad
	-
    \delta
    \big(
    \mathcal{E}_{\QQ}
	-
	\mathcal{E}_{\QQ'}
	-
	\hbar\omega_{|\QQ'-\QQ|}^\alpha
	\big)
	\Big)
	\bigg].
\end{align}
The inhomogeneous contribution determines the sign of $\delta N_\QQ$, i.e., bosonic (positive at $\QQ=0$) or fermionic (negative at $\QQ=0$) distributions compared to the classical case. We thus, after inserting Eq.~(\ref{eq:boltzmanndistribution}), focus on:

\begin{align}\label{eq:nsquareausgeklammert}
    &\partial_t
    (\delta    
    N_{\QQ}
    )
    \nonumber\\
    &\approx
	\frac{2\pi}{\hbar}
    \sum_{\alpha\QQ'}
    \big|
    g_{\QQ'-\QQ}^{c\alpha}
    -
    g_{\alpha,\QQ'-\QQ}^{v\alpha}
    \big|^2
    \pi^2\bar n^2\nonumber\\
    &\qquad
    \times
    \Big(
    (2\pi^\frac{1}{2}\lambda_{th})^{2d}
    e^{-\lambda_{th}^2K^2}
    e^{-\lambda_{th}^2Q^2}
    \nonumber\\
    &\qquad\qquad\qquad
    -
    2
    (2\pi^\frac{1}{2}\lambda_{th})^d
    a_0^d
    \mathcal{F}^d
    \big(
    e^{-\lambda_{th}^2K^2}
    +
    e^{-\lambda_{th}^2Q^2}
    \big)
    \nonumber\\
    &\qquad\qquad\qquad
    +
    (a_0^d)^2
    \mathcal{B}^d
    \Big)
    \nonumber\\
    &\qquad
    \times
    \Big(
    \delta
    \big(
    \mathcal{E}_{\QQ}
	-
	\mathcal{E}_{\QQ'}
	+
	\hbar\omega_{|\QQ-\QQ'|}^\alpha
	\big)
    \nonumber\\
    &\qquad\qquad\qquad
	-
    \delta
    \big(
    \mathcal{E}_{\QQ}
	-
	\mathcal{E}_{\QQ'}
	-
	\hbar\omega_{|\QQ'-\QQ|}^\alpha
	\big)
	\Big)
\end{align}

In Eq.~(\ref{eq:nsquareausgeklammert}), all three terms appear consistently in second order in $\bar n$. However, the first term, the ideal bosonic nonlinearity, is entirely independent of the Bohr radius $a_0$ and just depends on the thermal wavelength $\lambda_{th}$, i.e., will become important at low temperatures and for small excitonic masses, as one would expect for ideal bosons. The fermionic corrections, however, depend on the Bohr radius $a_0$. In order to predict the sign of the nonlinearity, Eq.~(\ref{eq:nsquareausgeklammert}), w.r.t.\ the ground state, we compute the equation for the occupation at $\QQ=0$, which allows us to eliminate the $\sum_{\QQ'}$ by the Fermi rule delta functions. This is in principle possible for arbitrary phonon dispersions, as long as one has an analytic expression for the phonon energy $\hbar\omega^\alpha$. To give an example, we illustrate it here for a monolayer TMDC, where typically, for optical phonons, one can assume e.g. zeroth order deformation potential, i.e., $\hbar\omega^{opt}\approx E_{opt}=const.$~\cite{berkelbach2013theory, selig2020suppression}, and thus $Q_\text{phon}^{opt}=\sqrt{\frac{2M}{\hbar^2}E_{opt}}$, while for acoustical phonons, one can typically assume first order deformation potential, thus a linear dispersion $\hbar\omega^{akk}\approx kc_{akk}$~\cite{berkelbach2013theory, selig2020suppression} and hence $Q_\text{phon}^{akk}=\frac{2M}{\hbar^2}c_{akk}$. This is however only to give an example, in other materials, different phonon modes will be important and hence the values of the momentum will also be different. In the following we will only assume that we obtained a $Q_\text{phon}^\alpha$ from the Dirac delta and eliminated the $\sum_{\QQ'}$ in Eq.~(\ref{eq:nsquareausgeklammert}). This gives
 \begin{widetext}

 \begin{align}\label{eq:finaleqafteranalyticderivation}
    \partial_t
    (
    \delta    
    N_{\QQ=0}    
    )
    &=
    \frac{L^d}{\hbar\pi^{d-2}}
    \sum_\alpha
    Q_\text{phon}^\alpha
    \big|
    g_{Q_\text{phon}^\alpha}^{c\alpha}
    -
    g_{Q_\text{phon}^\alpha}^{v\alpha}
    \big|^2
    \bar n^2
    \nonumber\\
    &
    \times
    \Big(
    (2\pi^\frac{1}{2}\lambda_{th})^{2d}
    e^{-\lambda_{th}^2(Q_\text{phon}^\alpha)^2}
    -
    2
    (2\pi^\frac{1}{2}\lambda_{th})^{d}
    a_0^d\mathcal{F}^d
    \big(    e^{-\lambda_{th}^2(Q_\text{phon}^\alpha)^2}
    +
    1
    \big)
    +
    (a_0^d)^2
    \mathcal{B}^d
    \Big)
\end{align}     

or, in 2d, i.e., for $d=2$
\begin{align}\label{eq:finiteeq2d}
    \partial_t
    (\delta    
    N_{\QQ=0}
    )
    &=
    \frac{L^2}{\hbar}
    \sum_\alpha
    Q_\text{phon}^\alpha
    \big|
    g_{Q_\text{phon}^\alpha}^{c\alpha}
    -
    g_{Q_\text{phon}^\alpha}^{v\alpha}
    \big|^2
    16\pi^2\bar n^2\nonumber\\
    &\times
    \underbrace{
    \Big(
    \lambda_{th}^4
    e^{-\lambda_{th}^2(Q_\text{phon}^\alpha)^2}
    -
    \frac{8}{5}\lambda_{th}^2a_0^2
    \big(
    e^{-\lambda_{th}^2(Q_\text{phon}^\alpha)^2}
    +
    1
    \big)
    +
    a_0^4
    \Big)}_
    {\equiv\lambda_{th}^4 f_{2d}^\alpha(\frac{a_0}{\lambda_{th}},\lambda_{th}Q_\text{phon}^\alpha)}
    .
\end{align}
and in 3d, i.e., $d=3$
\begin{align}\label{eq:finiteeq3d}
    \partial_t
    (\delta    
    N_{\QQ=0}
    )
    &=
    \frac{L^3}{\hbar}
    \sum_\alpha
    Q_\text{phon}^\alpha
    \big|
    g_{Q_\text{phon}^\alpha}^{c\alpha}
    -
    g_{Q_\text{phon}^\alpha}^{v\alpha}
    \big|^2
    16\pi^2\bar n^2\nonumber\\
    &
    \times
    \underbrace{
    \Big(
    64
    \lambda_{th}^6
    \pi
    e^{-\lambda_{th}^2(Q_\text{phon}^\alpha)^2}
    -
    264\sqrt{\pi}
    a_0^3\lambda_{th}^3
    \big(    
    e^{-\lambda_{th}^2(Q_\text{phon}^\alpha)^2}
    +
    1
    \big)
    +
    a_0^6
    \frac{4199}{8}
    \Big)}_
    {\equiv\lambda_{th}^6 f_{3d}^\alpha(\frac{a_0}{\lambda_{th}},\lambda_{th}Q_\text{phon}^\alpha)}.
\end{align}
 \end{widetext}
\section{Results in unitless parameters}\label{Sec:analyticresults}
\subsection{Individual phonon branches}
In Eq.~(\ref{eq:mainequation}), typically, both acoustic and optical phonon modes contribute to the index $\alpha$. In the following, we treat the phononic modes individually to determine the sign of the combined nonlinearities of each phononic mode on the energetically lowest excitonic state at $\QQ=0$. The border $f_{d}^\alpha(\frac{a_0}{\lambda_{th}},\lambda_{th}Q_\text{phon}^\alpha)=0$ between fermionic and bosonic behavior is determined by setting $\delta N_{\QQ=0}=0$: 
\begin{widetext}
    \begin{align}
    0&=
    \Big(
    (2\pi^\frac{1}{2}\lambda_{th})^{2d}
    e^{-\lambda_{th}^2(Q_\text{phon}^\alpha)^2}
    -
    2
    a_0^d\mathcal{F}^d
    (2\pi^\frac{1}{2}\lambda_{th})^{d}
    \big(    e^{-\lambda_{th}^2(Q_\text{phon}^\alpha)^2}
    +
    1
    \big)
    +
    (a_0^d)^2\mathcal{B}^d
    \big)
    \Big)
\end{align}
\end{widetext}

When divided by $(\lambda_{th}^{d})^2$, we can rewrite Eqs.~(\ref{eq:finiteeq2d},\ref{eq:finiteeq3d}) to:
\begin{align}\label{eq:coexcistence2dlam}
    &f_{2d}^\alpha(\frac{a_0}{\lambda_{th}},\lambda_{th}Q_\text{phon}^\alpha)
    \nonumber\\
    &=
    16
    e^{-\lambda_{th}^2(Q_\text{phon}^\alpha)^2}
    -
    \frac{32}{5} 
    \frac{a_0^2}{\lambda_{th}^2}
    \big(    
    e^{-\lambda_{th}^2(Q_\text{phon}^\alpha)^2}
    +
    1
    \big)
    +
    \frac{a_0^4}{\lambda_{th}^4}
\end{align}
and for $d=3$
\begin{align}\label{eq:coexcistence3dlam}
    &f_{3d}^\alpha(\frac{a_0}{\lambda_{th}},\lambda_{th}Q_\text{phon}^\alpha)
    \nonumber\\
    &=
    \pi
    e^{-\lambda_{th}^2(Q_\text{phon}^\alpha)^2}
    -
    \frac{33}{8}\sqrt{\pi}
    \frac{a_0^3}{\lambda_{th}^3}
    \big(    
    e^{-\lambda_{th}^2(Q_\text{phon}^\alpha)^2}
    +
    1
    \big)
    +
    \frac{a_0^6}{\lambda_{th}^6}
    \frac{4199}{512}
\end{align}
\begin{figure}[t!]
    \centering
    \includegraphics[width=\linewidth]{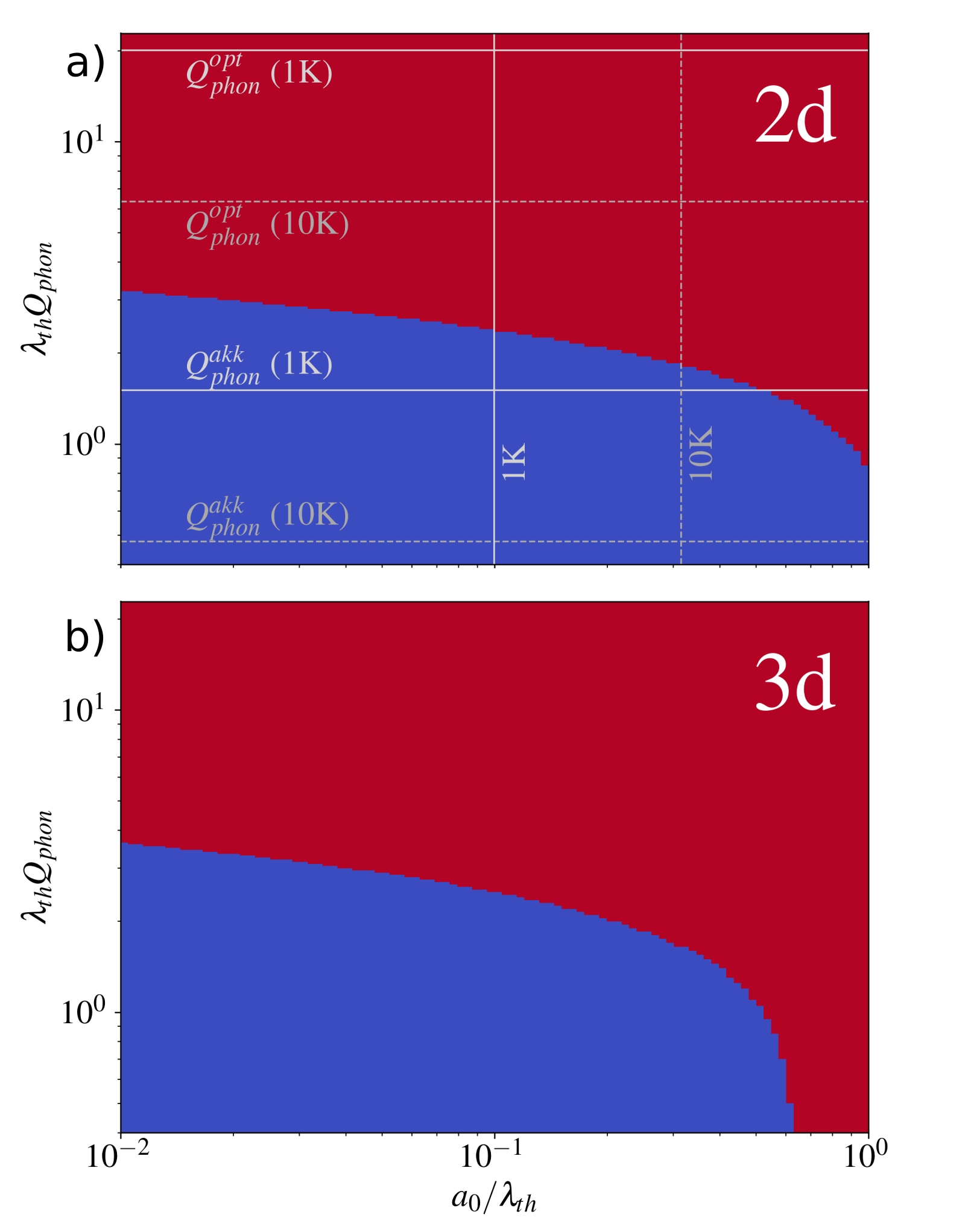}
    \caption{Plot of the sign of the nonlinearities deciding the thermalization occupation of the ground state, as a function of the unitless parameters $\frac{a_0}{\lambda_{th}}$ and $\lambda_{th}Q_\text{phon}^\alpha$. Blue stands for a positive sign of $f_d^\alpha(\frac{\lambda_{th}}{a_0},\lambda_{th}Q_\text{phon}^\alpha)$, indicating bosonic thermalization behavior, and red stands for a negative sign, respectively indicating fermionic behavior compared to the classical distribution.  a) 2d case, Eq.~(\ref{eq:coexcistence2dlam}). The gray lines in the 2d plot show temperatures and $\lambda_{th}Q_\text{phon}$ values for typical TMDC parameters, i.e., $a_0=2$nm and $M=1.1m_{el}$, in order to give an orientation in the parameter plane. b) Same plot for the 3d case, Eq.~(\ref{eq:coexcistence3dlam}). (Above $\frac{a_0}{\lambda_{th}}>1$ in principle there is another parameter range with positive sign, however, the approximations we made require $a_0 \ll \lambda_{th}$, and we thus only show the region that is in accordance with this approximation.)}
    \label{fig:approxlam}
\end{figure}
Interestingly, only the dimensionless parameters $\frac{a_0}{\lambda_{th}}$ and $\lambda_{th}Q_\text{phon}$ (or, as later shown, equivalently $a_0Q_\text{phon}$) determine the dynamics. Fig.~\ref{fig:approxlam} is a plot of the sign of Eqs.~(\ref{eq:coexcistence2dlam},\ref{eq:coexcistence3dlam}) over the unitless parameters $\frac{a_0}{\lambda_{th}}$ and $\lambda_{th}Q_\text{phon}$. This plot in principle applies in general for different semiconductor materials, as long as phonon scattering of two- or three dimensional semiconductor excitons dominates the dynamics. For both two and three dimensional excitons, one can see that as expected, the excitons become more bosonic with decreasing Bohr radius $a_0$, with 3d excitons being even more sensitive towards this radius, which makes sense as it enters the equation in powers of $a_0^d$, thus more dimensions $d$ make the dependency stronger. 

In App.~\ref{app:moreanalyticplots} we show that the limiting case of ideal bosonic behavior is included, which makes sense mathematically, when setting $a_0=0$ in Eqs.~(\ref{eq:coexcistence2dlam},\ref{eq:coexcistence3dlam}), only the first term prevails, which stands for bosonic stimulated scattering to the ground state similar to the behavior of ideal bosons. In the same appendix we also provide logplots showing that this limiting case is only approached very slowly, thus for all realistic Bohr radii we always have strong contributions from the fermionic corrections. 

In order to relate our findings to experimentally reasonable values for the unitless parameters, we exemplarily provide positions in this parameter plane for TMDC excitons, which we estimate to have a Bohr radius of $a_0=2$nm and a mass of $M=1.1 m_{el}$. This allows us to give exemplary lines for temperature (vertical) and for typical phonon - momenta $Q_\text{phon}$ (the momentum that a typical scattering event requires for scattering to the ground state).
It becomes evident from Fig.~\ref{fig:approxlam} that optical phonon processes require too large momenta to make bosonic behavior probable, while acoustical phonons are more likely to favor bosonic thermalization. The fact that the bosonic first term in Eqs.~(\ref{eq:coexcistence2dlam},\ref{eq:coexcistence3dlam}) only for very small scattering momenta dominates the second term, its fermionic counterpart, is here encoded in the exponential function we inserted for the occupation of the scattering partner $N_{\QQ_\text{phon}}$. Large momenta between the excitonic states for optical phonon scattering  lead to small occupation of the scattering partner, as already shown in Fig.~\ref{fig:boltzmann-sketch}. Inelastic, optical phonon scattering therefore hinders bosonic thermalization. This is in very good agreement with our findings from the full numerics~\cite{katzer2023excitonphonon}.

A shortcoming of the plot in Fig.~\ref{fig:approxlam} is that it is not very intuitive to read temperature dependencies from it. We therefore provide another set of equations, where $\lambda_{th}$ is only on one axis and thus the temperature dependence can be seen more directly. We will see that it does not alter our main finding that optical phonon modes (or, more general, inelastic phonon scattering) inhibits stimulated scattering effects:\\
\begin{figure}[t!]
    \centering
    \includegraphics[width=\linewidth]{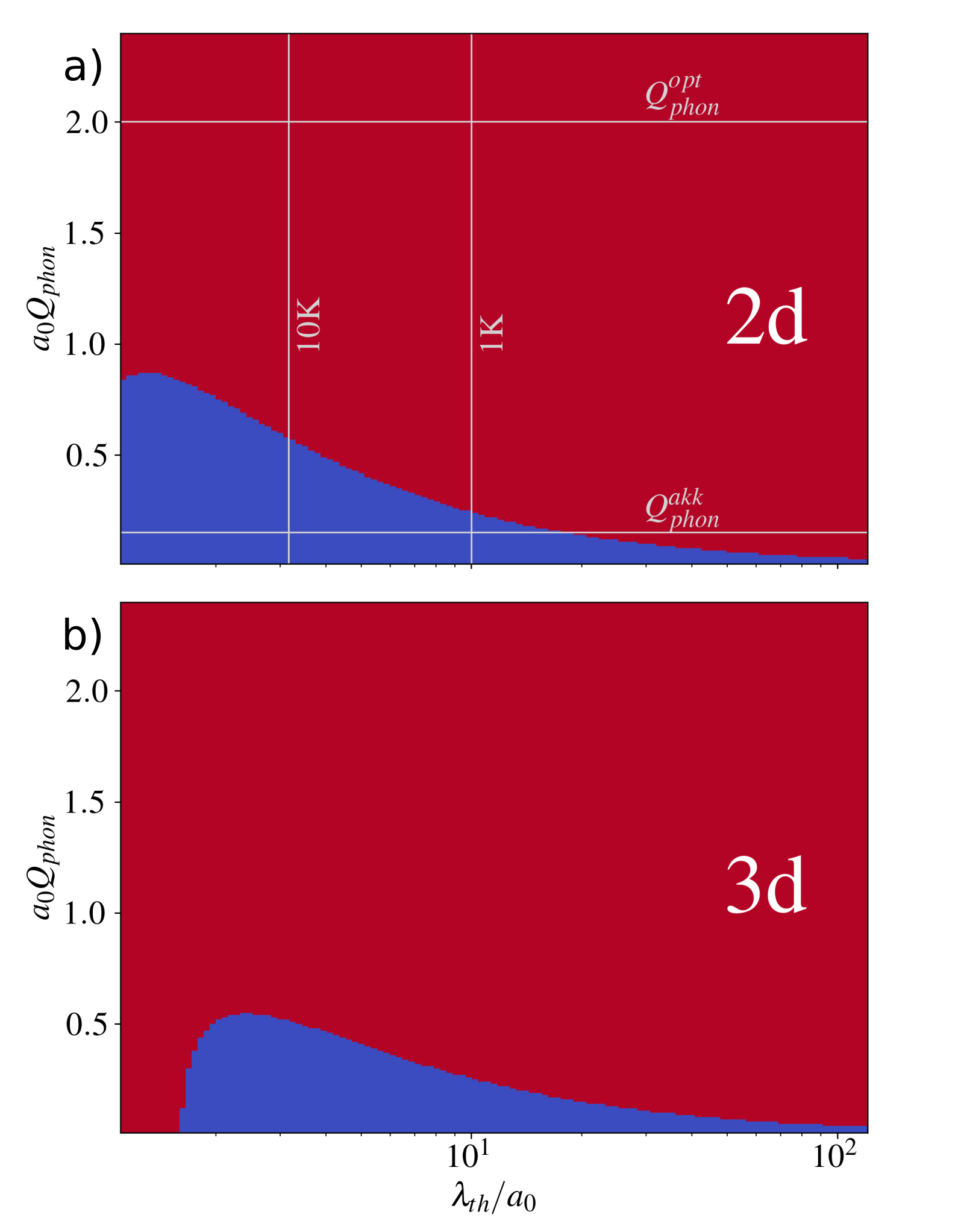}
    \caption{Plot of the sign of the nonlinearities deciding the thermalization occupation of the ground state, as a function of the unitless parameters $\frac{\lambda_{th}}{a_0}$ and $a_0Q_\text{phon}^\alpha$. Blue stands for a positive sign of $g_d^\alpha(\frac{\lambda_{th}}{a_0},a_0Q_\text{phon}^\alpha)$, indicating bosonic thermalization behavior, and red stands for a negative sign, respectively indicating fermionic behavior compared to the classical distribution.  a) 2d case, Eq.~(\ref{eq:coexcistence2d}). The gray lines show temperatures and $a_0Q_\text{phon}$ values for typical TMDC parameters for orientation, i.e., $a_0=2$nm and $M=1.1m_{el}$. b) Same plot for the 3d case, Eq.~(\ref{eq:coexcistence3d}). }
    \label{fig:approxa0}
\end{figure}
%
%
The whole expressions of Eqs.~(\ref{eq:finiteeq2d},\ref{eq:finiteeq3d}) divided by $(a_0^2)^d$ read for 2d
\begin{align}\label{eq:coexcistence2d}
    &g_{2d}^\alpha(\frac{\lambda_{th}}{a_0},a_0Q_\text{phon}^\alpha)\nonumber\\
    &=
    16\frac{\lambda_{th}^4}{a_0^4}
    e^{-\frac{\lambda_{th}^2}{a_0^2}(a_0Q_\text{phon}^\alpha)^2}
    \nonumber\\
    &\qquad\qquad\qquad\qquad
    -
    \frac{32}{5} 
    \frac{\lambda_{th}^2}{a_0^2}
    \big(    
    e^{-\frac{\lambda_{th}^2}{a_0^2}(a_0Q_\text{phon}^\alpha)^2}
    +
    1
    \big)
    +
    1
\end{align}
and for 3d
\begin{align}\label{eq:coexcistence3d}
    &g_{3d}^\alpha(\frac{\lambda_{th}}{a_0},a_0Q_\text{phon}^\alpha)
    \nonumber\\
    &=
    \pi\frac{\lambda_{th}^6}{a_0^6}
    e^{-\frac{\lambda_{th}^2}{a_0^2}(a_0Q_\text{phon}^\alpha)^2}
    \nonumber\\
    &\qquad
    -
    \frac{33}{8}\sqrt{\pi}
    \frac{\lambda_{th}^3}{a_0^3}
    \big(    
    e^{-\frac{\lambda_{th}^2}{a_0^2}(a_0Q_\text{phon}^\alpha)^2}
    +
    1
    \big)
    +
    \frac{4199}{512}.
\end{align}
(Note that we rewrote also the arguments of the expontentials in order to express everything w.r.t.\ the same unitless parameters.) 
Fig.~\ref{fig:approxa0} is a plot of the sign of Eqs.~(\ref{eq:coexcistence2d},\ref{eq:coexcistence3d}) as a function of the unitless parameters $\frac{\lambda_{th}}{a_0}$ and $a_0Q_\text{phon}^\alpha$. From the plot we can see that in order to expect bosonic signatures, the temperature has to be chosen low enough that for the respective particle mass, $\lambda_{th}$ can compensate the Bohr radius $a_0$. Furthermore, one requires values of $Q_\text{phon}^\alpha \ll \frac{1}{a_0}$ for 2d excitons or $Q_\text{phon}^\alpha \ll \frac{1}{2a_0}$ in the 3d case. Such low scattering momenta can typically only be provided by acoustic, quasielastic phonon scattering. For larger $Q_\text{phon}^\alpha$, the fermionic nonlinearity dominates (to be precise, the +1 next to the respective exponential function wins for too large values of $Q_\text{phon}^\alpha$ in Eq.~(\ref{eq:coexcistence2d})). Interestingly, 3d excitons are even more temperature-sensitive, with only a comparatively small window of temperatures apparently allowing for bosonic behavior, if the dominant phonon process is elastic enough.
For TMDC excitons, we can again estimate $a_0=2$nm and $M=1.1 m_{el}$ and give exemplary lines for temperature (vertical) and for typical phonon - momenta $Q_\text{phon}^\alpha$ (the momentum that a typical scattering event requires for scattering to the ground state). For the example of TMDC excitons, temperatures between $T=\unit[1-10]{K}$ look promising for acoustic phonon branches to give positive values, i.e., dominant bosonic signatures. As discussed earlier, optical phonon scattering to the ground state requires much larger momenta, since those scattering processes are much less elastic. The $Q_\text{phon}^{opt}$ are thus typically one order of magnitude larger than for the acoustic branches, at least in the monolayer~\cite{li2013intrinsic,jin2014intrinsic}. They will thus contribute with a negative sign, and lead to a fermionic thermalization behavior. 

\subsection{Sum over phonon branches}

In the more realistic case of taking all phonon branches into account that are relevantly contributing to the scattering process, we have to consider the sum over $\alpha$ in Eq.~(\ref{eq:finaleqafteranalyticderivation}), and thus get prefactors for the contributions from the different branches, which depend on the $Q_\text{phon}^\alpha$ of the respective phonon mode $\alpha$, which can be written as
\begin{align}
    &h_d(\frac{\lambda_{th}}{a_B},\lambda_{th})
    \nonumber\\
    &=
    \sum_\alpha
    Q_\text{phon}^\alpha
    \big|
    g_{Q_\text{phon}^\alpha}^{c\alpha}
    -
    g_{Q_\text{phon}^\alpha}^{v\alpha}
    \big|^2
    f_d^\alpha
    (\frac{\lambda_{th}}{a_B},\lambda_{th}Q_\text{phon}^\alpha),
\end{align}
with, for the example of TMDC excitons:
\begin{align}
    &Q_\text{phon}^\text{opt}|g_{Q_\text{phon}}^{c,\text{opt}}-g_{Q_\text{phon}}^{v,\text{opt}}|^2
    \approx
    2000\,
    Q_\text{phon}^\text{akk}|g_{Q_\text{phon}}^{c,\text{akk}}-g_{Q_\text{phon}}^{v,\text{akk}}|^2
\end{align}
Due to the mentioned larger values for $Q_\text{phon}$ and the fact, that $g_{Q_\text{phon}^{c,v}}$ is significantly larger for the optical phonons, the prefactor $Q_\text{phon}|g_{Q_\text{phon}}^{c\alpha}-g_{Q_\text{phon}}^{v\alpha}|^2$ for instance is around 2000 times larger than for the acoustical branch in a TMDC monolayer~\cite{li2013intrinsic,jin2014intrinsic,selig2016excitonic}. This also applies for the linear equation, however, there the acoustic phonons become important once the $n_{\QQ'}^\alpha\gg 1$, while the optical mode freezes out~\cite{selig2018dark,selig2019ultrafast,selig2020suppression}. However, the nonlinear terms are not directly dependent on temperature, thus for the nonlinearity, optical phonon modes will probably always dominate, at least in monolayer TMDCs.

Van der Waals heterostructures of TMDCs and excitons in other kinds of semiconductors may show different behavior, if the optical phonon modes are absent, or less dominant. Our study suggests that the macroscopic occupation of the lowest state becomes more probable in systems with dominating acoustic phonons. This is supported by the full numerical solution of Eq.~(\ref{eq:mainequation}): If we simulate the thermalization for only acoustical phonon branches, the thermalization shows a bosonic behavior for much larger Bohr radii, far beyond the TMDC limit of $a_0 = \unit[2]{nm}$, as can be seen in Fig.~2(b) in Ref.~\cite{katzer2023excitonphonon}.

\section{Conclusion}\label{Sec:Conclusion}
We analytically discussed a recently derived equation for the exciton phonon kinetics~\cite{katzer2023excitonphonon} above the linear zero density limit. The kinetic equation is microscopically derived from the electron hole picture, taking the next order in $\eta=\bar na_B^2$ into account, thus going beyond the bosonic commutator relation for those composite particles. In a fully analytic approach we discussed the effect of Bohr radius, thermal wavelength and typical phonon scattering momentum on the ground state occupation, to study the question whether the overall thermalization can be considered bosonic or fermionic, and make general statements in a framework of unitless parameters, such as $\frac{a_0}{\lambda_{th}}$, $\lambda_{th}Q_{phon}$ and $a_0Q_{phon}$. Conducting the derivation in a dimension-independent approach allowed us to give predictions for both 2d and 3d exciton systems.
As demonstrated before in our numerical study~\cite{katzer2023excitonphonon}, also in the analytical limit we show drastic deviations from a purely bosonic behavior, and show that for typical Bohr radii of around $a_0=\unit[2]{nm}$ for TMDC excitons, the compound particles cannot be considered bosonic and thus are not likely to show macroscopic occupation effects for the ground state, as long as optical phonon scattering dominates the thermalization dynamics. For significantly smaller Bohr radii, such as for instance the reported $a_0\approx\unit[0.6]{nm}$ for the antiferromagnet van der Waals material NiPS$_3$~\cite{dirnberger2022spincorrelated}, or with absent optical phonon modes at low temperatures, we showed that the respective excitons can be expected to show a bosonic behavior, as the bosonic stimulated scattering in this regime would overcompensate the weaker fermionic Pauli blocking. 
\section{ACKNOWLEDGMENTS}
We thank Dominik Christiansen, Emil Denning, Aycke Roos and Marten Richter from TU Berlin, 
Mirco Troue, Johannes Figueiredo, Lukas Sigl and Alexander Holleitner from TU München and Ursula Wurstbauer from University of Münster
for fruitful discussions, and gratefully acknowledge support from the Deutsche Forschungsgemeinschaft (DFG) through SFB 951, project number 182087777, and via grant KN 427/11-2, project number 420760124.

\appendix
\section{Wavefunction overlaps}\label{app:overlapappendix}
The wavefunction overlaps for the fermionic tensors, Eq.~(\ref{eq:fermionicscatteringtensor}), read
\begin{widetext}
\begin{align}\label{eq:fermionicoverlaps}
    g_{\QQ'\QQ,\KK,\alpha}^{F,\lambda\nu,\nu'}
    &=
    \sum_\qq
    g_{\QQ'-\QQ}^{c\alpha}
    \Big(
    (\varphi_{\qq}^{\lambda})^*
    (\varphi_{\qq+\tilde\beta\KK+\tilde\alpha\QQ'-\QQ}^{\nu'})^*
    \varphi_{\qq+\tilde\beta(\KK-\QQ')}^{\nu'}
    \varphi_{\qq-\tilde\alpha(\QQ-\QQ')}^{\nu}
    +
    (\varphi_{\qq}^{\lambda})^*
    (\varphi_{\qq-\tilde\alpha(\KK-\QQ')}^{\nu'})^*
    \varphi_{\qq-\tilde\alpha(\KK-\QQ')}^{\nu'}
    \varphi_{\qq+\tilde\beta(\QQ-\QQ')}^{\nu}
    \Big)
    \nonumber\\
    &\qquad
    -
    g_{\QQ'-\QQ}^{v\alpha}
    \Big(
    (\varphi_{\qq}^{\lambda})^*
    (\varphi_{\qq+\tilde\beta(\KK-\QQ')}^{\nu'})^*
    \varphi_{\qq+\tilde\beta(\KK-\QQ')}^{\nu'}
    \varphi_{\qq-\tilde\alpha(\QQ-\QQ')}^{\nu}
    +
    (\varphi_{\qq}^{\lambda})^*
    (\varphi_{\qq-\tilde\alpha\KK-\tilde\beta\QQ'+\QQ)}^{\nu'})^*
    \varphi_{\qq-\tilde\alpha(\KK-\QQ')}^{\nu'}
    \varphi_{\qq+\tilde\beta(\QQ-\QQ')}^{\nu}
    \Big),
\end{align}
and for the exchange tensor, Eq.~(\ref{eq:exchangescatteringtensor}),
\begin{align}\label{eq:exchangeoverlaps}
    g_{\QQ\QQ',\KK,\KK',\alpha}^{E,\lambda\nu\lambda'\nu'}
    &=
    \sum_\qq
    \Big(
    g_{\QQ-\QQ'}^{c\alpha}
    (\varphi_{\qq}^{\nu})^*
    (\varphi_{\qq+\tilde\alpha(\QQ-\KK')}^{\lambda'})^*
    (\varphi_{\qq+\tilde\beta\KK+\tilde\alpha\QQ-\QQ'}^{\nu'})^*
    \varphi_{\qq+\tilde\beta\KK-\tilde\beta\QQ}^{\nu'}
    \varphi_{\qq+\tilde\alpha(\QQ-\KK')}^{\lambda'}
    \varphi_{\qq+\tilde\alpha(\QQ-\QQ')}^{\lambda}\nonumber\\
    &\qquad\qquad
    +
    g_{\QQ-\QQ'}^{c\alpha}
    (\varphi_{\qq}^{\nu})^*
    (\varphi_{\qq-\KK+\tilde\beta\KK'+\tilde\alpha\QQ}^{\lambda'})^*
    (\varphi_{\qq-\tilde\alpha\KK+\KK'+\tilde\alpha\QQ-\QQ'}^{\nu'})^*
    \varphi_{\qq+\tilde\beta\KK'-\tilde\beta\QQ}^{\lambda'}
    \varphi_{\qq-\tilde\alpha(\KK-\QQ)}^{\nu'}
    \varphi_{\qq-\KK+\KK'+\tilde\alpha(\QQ-\QQ')}^{\lambda}\nonumber\\
    %
    &\qquad\qquad
    -
    g_{\QQ-\QQ'}^{v\alpha}
    (\varphi_{\qq}^{\nu})^*
    (\varphi_{\qq-\tilde\alpha\KK'-\tilde\beta\QQ+\QQ'}^{\lambda'})^*
    (\varphi_{\qq+\tilde\beta(\KK-\QQ)}^{\nu'})^*
    \varphi_{\qq+\tilde\beta(\KK-\QQ)}^{\nu'}
    \varphi_{\qq-\tilde\alpha\KK'+\tilde\alpha\QQ}^{\lambda'}
    \varphi_{\qq-\tilde\beta(\QQ-\QQ')}^{\lambda}\nonumber\\
    &\qquad\qquad
    -
    g_{\QQ-\QQ'}^{v\alpha}
    (\varphi_{\qq}^{\nu})^*
    (\varphi_{\qq-\KK+\tilde\beta\KK'-\tilde\beta\QQ+\QQ'}^{\lambda'})^*
    (\varphi_{\qq-\tilde\alpha\KK+\KK'-\tilde\beta\QQ}^{\nu'})^*
    \varphi_{\qq-\tilde\beta(\QQ-\KK')}^{\lambda'}
    \varphi_{\qq-\tilde\alpha\KK+\tilde\alpha\QQ}^{\nu'}
    \varphi_{\qq-\KK+\KK'-\tilde\beta(\QQ-\QQ')}^{\lambda}
    \Big).
\end{align}
\end{widetext}
\section{Integrals over wavefunctions}\label{app:wavefunctionintegrals}
In Sec.~\ref{sec:lowtempsubsection} we introduced abbreviations, which essentially give the result of integrals over not only two $\sum_\qq |\varphi_q|^2=1$, but four or six wavefunctions, to approximate the integrals of Eq.~(\ref{eq:fermionicoverlaps},\ref{eq:exchangeoverlaps}). In 2d this reads:
\begin{align}
	\mathcal{F}^{d=2}
	=
	&\frac{L^2}{a_0^2}\sum_\qq |\varphi_q|^4 
	= 
	\frac{4\pi}{5} 
	\\
	\mathcal{B}^{d=2}
	=
	&\frac{L^4}{a_0^4}\sum_\qq |\varphi_q|^6 
	=  
	\pi^2
\end{align}
and in 3d:
\begin{align}    
	\mathcal{F}^{d=3}
	=
	&\frac{L^3}{a_0^3}\sum_\qq |\varphi_q|^4 
	= 
	\frac{33\pi}{2} 
	\\
	\mathcal{B}^{d=3}
	=
	&\frac{L^6}{a_0^6}\sum_\qq |\varphi_q|^6 
	=  
	\frac{4199\pi^2 }{8}
\end{align}

%


\section{The limit of small Bohr radii}\label{app:moreanalyticplots}

\begin{figure}[h!]
    \centering
    \includegraphics[width=\linewidth]{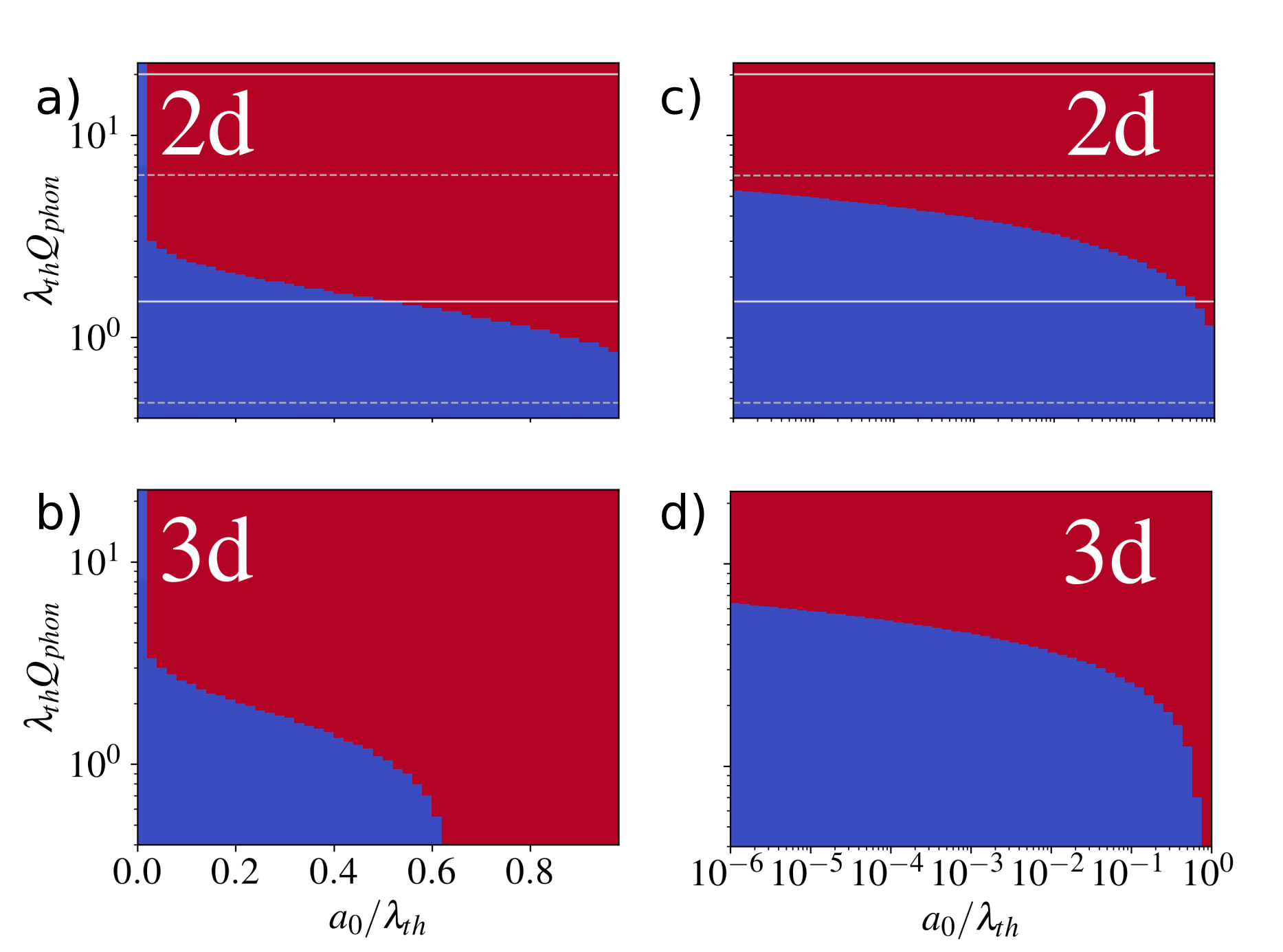}
    \caption{Plot of the sign of Eqs.~(\ref{eq:coexcistence2dlam},\ref{eq:coexcistence3dlam}).   Blue stands for a positive sign of $f(a_0/\lambda_{th},\lambda_{th}Q_\text{phon})$, and red for a negative one, respectively. The gray lines for the 2d plots show temperature dependent $\lambda_{th}Q_\text{phon}$ values for TMDC-like values, i.e., $a_0=2$nm and $M=1.1m_{el}$, analogous to Fig.~\ref{fig:approxlam}. The main point here is that the theoretical limit of $a_0=0$ gives ideal bosons that show stimulated scattering independent of the phonon scattering momenta. This limit is however not experimentally accessible, as even for arbitrarily small Bohr radii the fermionic corrections remain important.}
    \label{fig:a0limitplots}
\end{figure}

Our theory comprises the ideal case of pure bosonic particles for vanishing Bohr radius $a_0=0$, where e.g. in Eqs.~(\ref{eq:coexcistence2dlam},\ref{eq:coexcistence3dlam}), only the nonlinearity for bosonic stimulated scattering prevails, and the bosonic character occurs independent of thermal wavelength (and thus of temperature and particle mass) and independent of the details of the phonon coupling. This limit is shown in Figs.~\ref{fig:a0limitplots}(a,b) respectively. However, this limit is of pure theoretical nature; even for arbitrarily small but finite values of $a_0$, the equation remains highly sensitive towards the phonon scattering momentum, as can be seen in the logplot, Figs.~\ref{fig:a0limitplots}(c,d). Even for arbitrarily small Bohr radii the fermionic corrections remain important.
\section{Exemplary material parameters for MoSe$_2$}\label{app:parameters}
For details on the implementation of the parameters, see also~\cite{selig2016excitonic,selig2018dark,selig2019ultrafast,selig2020suppression,katzer2023impactnessy,katzer2023excitonphonon}.\\

\begin{table}[H]
\centering
  \begin{tabular}{c|l}
$e$ & \unit[1]{eC}\\
$c$ & \unit[299.7925]{nm/fs}\\
$\hbar$ & \unit[0.658212196]{eVfs} \\
$k_B$ & \unit[8.61745]{$\times$10$^{-5}$eV/K} \\
$\epsilon_0$ & \unit[5.526308]{$\times$10$^{-2}$ eC$^2$/(eV nm)} \\
$\mu_0$ & \unit[2.013384742]{$\times$10$^{-4}$ eV fs$^2$/(eC$^2$ nm)} \\
$m_{el}$ & \unit[5.6856800]{fs$^2$ eV/nm$^2$}  \\
$m_{P}$ & \unit[10439.60413]{fs$^2$ eV/nm$^2$}  \\
 \end{tabular}

\caption{\text{Important constants in semiconductor units.}}
\end{table}

\begin{table}[H]
\centering
  \begin{tabular}{c|c||c|c}

$c^{LA}$/10$^{-3}$nm fs$^{-1}$   & 4.1 & $c^{TA}$/10$^{-3}$nm fs$^{-1}$  & 4.1 \\
\hline
 $\hbar \omega^{\Gamma\,A'}$/meV  & 30.3
 & $\hbar \omega^{\Gamma\,TO}$/meV  & 36.1 

 \end{tabular}\label{tab_phon_dis}
  \caption{\textit{Phonon Dispersion.} Velocity of sound for the acoustic long range modes $c^{i}$ and phonon energies $\hbar \omega^{i}$ for optical modes, taken from \cite{jin2014intrinsic}.}
\end{table}

\begin{table}[H]
\centering
 \begin{tabular}{c|cc}
    
$a_0$/nm  & 0.3319 & \cite{rasmussen2015computational} 
\\
$d_0$/nm  & 0.34371 & \cite{rasmussen2015computational} 
\\
$\epsilon_\perp$  & 15.27  & \cite{berkelbach2013theory}
\end{tabular}\label{table_lattice}
 \caption{\textit{General Material Parameters.} We give the lattice constant $a_0$ and the distance between the two selenium atoms $d_0$. Additionally we require the inplane component of the respective dielectric tensor.}
\end{table}

\begin{table}[H]
\centering
\begin{tabular}{c|c}
$m_{eK}^\uparrow/m_{el}$  & 0.50
\\
$m_{hK}^\uparrow/m_{el}$ & 0.60 
 \end{tabular}\label{tab_el_dis}
 \caption{\textit{Effective masses} 
  taken from first principle computations (PBE) \cite{kormanyos2015theory}. 
 }
\end{table}

\begin{table}[H]
\centering
 \begin{tabular}{c||c|c||c|c}
 Trans. (Momentum) & conduction band & & valence band &  \\
   \hline
 $K\rightarrow K$ ($\Gamma$) & $D^{a}_1$/eV & 3.4  & $D^{a}_1$/eV  & 2.8 \\
 
  & $D^{o}_0$/eV nm$^{-1}$ & 52 & $D^{o}_0$/eV nm$^{-1}$ & 49 
  \\

\end{tabular}\label{tab_e_phon}
 \caption{
 \textit{Electron Phonon Coupling.} 
 Electron phonon coupling parameters in effective deformation potential approximation. The electron phonon matrix element is then given by $g^{i} = \sqrt{\frac{\hbar}{2 \rho \omega^i A}} V_q$, with $\rho$ being the mass density of the unit cell and $A$ being the semiconductor area (which cancels for all calculations). in the case of acoustic long range phonons, the coupling is given by the first order deformation potential $V_q=D_1 q$, whereas in the case of optical phonons and zone edge phonons, the coupling is given by zeroth order deformation potential coupling $V_q=D_0$. The parameters were taken from \cite{li2013intrinsic,jin2014intrinsic}.}
\end{table}


 


\bibliographystyle{apsrev4-1}
\bibliography{bib}
\end{document}